\documentclass[twocolumn,aps,prr,showpacs,longbibliography,superscriptaddress]{revtex4-1}
\usepackage{amssymb,latexsym}
\usepackage{mathrsfs}
\usepackage{lipsum}
\usepackage{graphicx}
\usepackage[caption=false]{subfig}
\usepackage{mathtools}
\usepackage{epstopdf}
\usepackage{xcolor}
\usepackage[colorlinks, linkcolor=red, anchorcolor=green, citecolor=green, urlcolor=blue]{hyperref}
\DeclareGraphicsExtensions{.pdf,.jpg,.png,.eps}
\usepackage[toc,page,title,titletoc,header]{appendix}
\usepackage{etoolbox}
\usepackage{breqn}

\makeatletter
\let\cat@comma@active\@empty
\makeatother

\begin{document}
	
\title{Precision bounds for quantum phase estimation using two-mode squeezed Gaussian states}

\author{Jian-Dong Zhang}
\email[]{zhangjiandong@jsut.edu.cn}
\affiliation{School of Mathematics and Physics, Jiangsu University of Technology, Changzhou 213001, China}
\author{Chuang Li}
\affiliation{Research Center for Novel Computing Sensing and Intelligent Processing, Zhejiang Lab, Hangzhou 311121, China}
\author{Lili Hou}
\affiliation{School of Mathematics and Physics, Jiangsu University of Technology, Changzhou 213001, China}
\author{Shuai Wang}
\affiliation{School of Mathematics and Physics, Jiangsu University of Technology, Changzhou 213001, China}

\date{\today}
	
\begin{abstract}
Quantum phase estimation based on Gaussian states plays a crucial role in many application fields. 
In this paper, we study the precision bound for the scheme using two-mode squeezed Gaussian states.
The quantum Fisher information is calculated and its maximization is used to determine the optimal parameters.
We find that two single-mode squeezed vacuum states are the optimal inputs and the corresponding precision bound is superior to the Heisenberg limit by a factor of 2.
For practical purposes, we consider the effects originating from photon loss.
The precision bound can still outperform the shot-noise limit when the lossy rate is below 0.4.
Our work may demonstrate a significant and promising step towards practical quantum metrology.
\end{abstract}

\maketitle

\section{Introduction}
Phase estimation based on optical interferometers is a fundamental means to achieve high-precision measurements for many important physical quantities, concentration, magnetic fields, gravitational waves, to name a few.
Quantum phase estimation can provide enhanced precision beyond the shot-noise limit, which is the precision bound attainable by exploiting classical resources.
In this regard, the quantum Fisher information is an effective tool to evaluate the precision bound for a specific input and parameterization \cite{PhysRevLett.72.3439,Liu_2020}.
It is particularly important for single-phase estimation, for the precision bound given by the quantum Fisher information can always be asymptotically saturated through specific positive-operator-valued measure (POVM) and maximum likelihood estimation.
The quantum Fisher information corresponding to the shot-noise limit and the Heisenberg limit can be expressed as ${\cal F}_{\rm {SNL}} = N$ and ${\cal F}_{\rm {HL}} = N^2$, where $N$ is the total average photon number employed for phase estimation.

In terms of the parameter estimation theory, any interferometer can be divided into three parts including probe preparation, phase encoding and POVM. 
The precision bound is completely determined by the first two parts. 
In general, the part of phase encoding is deterministic; therefore, improving the precision bound requires the engineering of the probe preparation. 
Related to this, two methods are usually deployed. 
The first method utilizes a linear beam splitter to combine two single-mode inputs, while the two inputs in the second method are combined by a nonlinear beam splitter, i.e., optical parametric amplifier (OPA).
The interferometers with these two methods are known as linear and nonlinear interferometers, respectively.

Over the past decades, numerous efforts have been made in quantum phase estimation based on linear interferometers.
Many exotic quantum states were considered as the input, such as two-mode squeezed vacuum states \cite{PhysRevLett.104.103602}, N00N states \cite{07510802091298}, entangled coherent states \cite{PhysRevLett.107.083601}, twin Fock states \cite{PhysRevA.68.023810} and coherent along with squeezed vacuum states \cite{PhysRevLett.100.073601}.
The precision bounds for the schemes using these above states can outperform the shot-noise limit and reach the Heisenberg limit.
In recent years, quantum phase estimation using nonlinear interferometers has also received lots of attention \cite{Plick_2010,PhysRevA.85.023815,PhysRevA.86.023844,PhysRevA.93.023810,PhysRevA.94.063840,Chekhova:16,10.1063/5.0004873}.
In this configuration, two inputs undergo a two-mode squeezing process provided by the first OPA. 
The correlation between the two modes are improved, and the total average number of photons is also increased.
These two types of interferometers have their own advantages. 
As a result, quantum phase estimation using hybrid or nested interferometers composed of nonlinear and linear beam splitters has also drawn considerable interest \cite{PhysRevLett.124.173602,PhysRevLett.128.033601,PhysRevA.87.023825,Zhang:18,PhysRevA.103.032617}.

The aforementioned studies analyzed the precision bounds for the schemes using some specific inputs. 
Recently, exploring the optimal inputs by maximizing the quantum Fisher information has been reported in a linear interferometer.
Lang \emph{et al.} analyzed the optimal input for the second port with the first port fed by a coherent state
\cite{PhysRevLett.111.173601}.
Zhang \emph{et al.} discussed the optimal single-mode input \cite{Zhang:22}.
The optimal separable Gaussian inputs were showed by Sparaciari \emph{et al.} \cite{Sparaciari:15}.
In this paper, we extend the problem of determining the optimal inputs to nonlinear interferometers.
We consider two general single-mode Gaussian states as the inputs.
The precision bound given by the quantum Fisher information is calculated and maximized by selecting the best parameters.
We analyze the precision bound in a lossy environment and discuss the tolerance against photon loss.
This work may provide a positive complement to the aspect of quantum phase estimation using nonlinear interferometers and relevant variants.

The remainder of this paper is organized as follows.
Section \ref{II} introduces the estimation scheme and provides the general expression for the quantum Fisher information.
Section \ref{III} determines the optimal inputs by maximizing the quantum Fisher information, and the corresponding precision bound is analyzed.
In Sec. \ref{IV}, we study the precision bound in the presence of photon loss and compare it with the precision bound of a classical-quantum hybrid scheme.
Finally, we summarize main results in Sec. \ref{V}.

\section{Estimation scheme and quantum Fisher information}
\label{II}

\begin{figure}[htbp]
	\centering
	\includegraphics[width=0.4\textwidth]{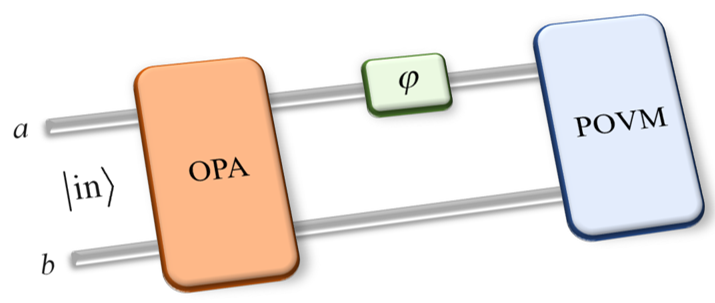}
	\caption{ Schematic diagram of quantum phase estimation using two-mode squeezed Gaussian states. OPA, optical parametric amplifier; POVM, positive-operator-valued measure.}
	\label{system}
\end{figure}

Figure \ref{system} gives the schematic diagram of quantum phase estimation using two-mode squeezed Gaussian states.
Two single-mode Gaussian states pass through an OPA and evolve into two-mode squeezed Gaussian states.
The estimated phase $\varphi$ is encoded into the state in mode $a$, and a POVM is performed.

Since the two-mode squeezing process of OPA is fixed, the task in this paper is to find the optimal Gaussian inputs.
The most general single-mode Gaussian states can be expressed as displaced squeezed thermal states \cite{RevModPhys.84.621}.
Meanwhile, it is no use to prepare thermal states as the inputs \cite{PhysRevLett.98.160401}. 
As a consequence, we only need to consider pure Gaussian inputs, two single-mode displaced squeezed vacuum states.
Without loss of generality, we assume that two states in modes $a$ and $b$ are the same.
Specifically, the input can be written as 
\begin{align}
\left| {{\rm{in}}} \right\rangle  = D\left( \alpha  \right)S\left( \xi  \right)\left| 0 \right\rangle  \otimes D\left( \alpha  \right)S\left( \xi  \right)\left| 0 \right\rangle, 	
\label{}
\end{align}
where $D\left( \alpha  \right)$ is the displacement operator with $\alpha  = \left| \alpha  \right|{e^{i\delta }}$ being the displacement amplitude, $S\left( \xi  \right)$ is the squeezing operator with
$\xi  = r{e^{i\theta }}$ being the squeezing parameter.
Further, the average photon number of the inputs is given by
\begin{align}
{N_{{\rm{in}}}} = \left\langle {{\rm{in}}} \right|\left( {{a^\dag }a + {b^\dag }b} \right)\left| {{\rm{in}}} \right\rangle  = {\rm{2}}( {{{\left| \alpha  \right|}^{\rm{2}}} + {{\sinh }^2}r} ),	
	\label{}
\end{align}
where $a^\dag$ ($b^\dag$) and $a$ ($b$) are creation and annihilation operators for mode $a$ ($b$).
The relation between the mode operators and displacement operator is
\begin{align}
{D^\dag }\left( \alpha  \right)aD\left( \alpha  \right) &= a + \alpha, \\
{D^\dag }\left( \alpha  \right)bD\left( \alpha  \right) &= b + \alpha, 
\label{}
\end{align}
and the relation between the mode operators and single-mode squeezing operator is
\begin{align}
{S^\dag }\left( \xi  \right)aS\left( \xi  \right) &= a\cosh r - {e^{i\theta }}{a^\dag }\sinh r,\\
{S^\dag }\left( \xi  \right)bS\left( \xi  \right) &= b\cosh r - {e^{i\theta }}{b^\dag }\sinh r.	
	\label{}
\end{align}

Due to the gain of the OPA, the total average number of photons in our scheme is found to be
\begin{align}
\nonumber N &= \left\langle {{\rm{in}}} \right|S_{\rm {OPA}}^\dag \left( g \right)( {{a^\dag }a + {b^\dag }b} )S_{\rm {OPA}}^{}\left( g \right)\left| {{\rm{in}}} \right\rangle  \\
&= {N_{{\rm{in}}}}\cosh {\rm{2}}g + 2{\left| \alpha  \right|^{\rm{2}}}\cos 2\delta \sinh 2g + {\rm{2}}{\sinh ^{\rm{2}}}g,
	\label{}
\end{align}
where we used the following relation
\begin{align}
S_{\rm {OPA}}^\dag \left( g \right)aS_{\rm {OPA}}^{}\left( g \right) &= a\cosh g + {b^\dag }\sinh g\\
S_{\rm {OPA}}^\dag \left( g \right)bS_{\rm {OPA}}^{}\left( g \right) &= b\cosh g + {a^\dag }\sinh g
	\label{}
\end{align}
with $S_{\rm {OPA}}^{}\left( g \right)$ being the two-mode 
squeezing operator and $g$ being the gain.

Now we analyze the precision bound of our scheme.
Since the inputs and operations are Gaussian, all information regarding the estimated phase can be obtained via the mean (first moment) and variance  (second moment) of the outputs.
For this reason, we calculate the quantum Fisher information through the use of symplectic geometry method.
Let us consider the vector composed of quadrature operators of modes $a$ and $b$,
\begin{align}
{\mathbf{X}} = [ {\begin{array}{*{20}{c}}
	{{x_a}}&{{p_a}}&{{x_b}}&{{p_b}}
	\end{array}} ]^{\mathsf{T}},
\label{}
\end{align}
where
\begin{align}
{x_m} &= {m^\dag } + m\\
{p_m} &= i( {{m^\dag } - m} )
\label{}
\end{align}
with $m \in \left\{ {a,b} \right\}$.
Then the mean vector of the inputs is given by 
\begin{align}
{\mathbf{M}_{\rm{in}}} = \left\langle {\mathbf{X}} \right\rangle  = 2 | \alpha | \cdot [ {\begin{array}{*{20}{c}}
	{\cos \delta }&{\sin \delta }&{\cos \delta }&{\sin \delta }
	\end{array}} ]^{\mathsf{T}},
\label{}
\end{align}
and we can find the covariance matrix of the inputs
\begin{align}
{\mathbf{\Sigma} _{{\rm{in}}}} = \left[ {\begin{array}{*{20}{c}}
	{{e^{2r}}}&0&0&0\\
	0&{{e^{ - 2r}}}&0&0\\
	0&0&{{e^{2r}}}&0\\
	0&0&0&{{e^{ - 2r}}}
	\end{array}} \right],
\label{}
\end{align}
whose arbitrary matrix element is defined by
\begin{align}
{\mathbf{\Sigma} _{kn}} = \frac{1}{2}\left\langle {{{\rm{X}}_k}{{\rm{X}}_n} + {{\rm{X}}_n}{{\rm{X}}_k}} \right\rangle  - \left\langle {{{\rm{X}}_k}} \right\rangle \left\langle {{{\rm{X}}_n}} \right\rangle. 
\label{}
\end{align}

Based on the relations between optical operations and quadrature operators, we can write the transformation matrix for OPA
\begin{align}
{{\mathbf{U}}_{{\rm{OPA}}}} = \left[ {\begin{array}{*{20}{c}}
	{\cosh g}&0&{\sinh g}&0\\
	0&{\cosh g}&0&{ - \sinh g}\\
	{\sinh g}&0&{\cosh g}&0\\
	0&{ - \sinh g}&0&{\cosh g}
	\end{array}} \right]
\label{}
\end{align}
and that for phase encoding
\begin{align}
{{\mathbf{U}}_{{\rm{PE}}}} = \left[ {\begin{array}{*{20}{c}}
	{\cos \varphi }&{ - \sin \varphi }&0&0\\
	{\sin \varphi }&{\cos \varphi }&0&0\\
	0&0&1&0\\
	0&0&0&1
	\end{array}} \right].
\label{}
\end{align}
By using the following transformations 
\begin{align}
{\mathbf{M}_{\varphi}} = {{\mathbf{U}}_{{\rm{PE}}}}{{\mathbf{U}}_{{\rm{OPA}}}}{{\mathbf{M}}_{{\rm{in}}}},
\label{}
\end{align}
\begin{align}
{\mathbf{\Sigma}_{\varphi}} = {\mathbf{U}}_{\rm{PE}}^{}{\mathbf{U}}_{\rm{OPA}}^{}\Sigma _{\rm{in}}^{}{\mathbf{U}}_{\rm{OPA}}^{\mathsf{T}} {\mathbf{U}}_{\rm{PE}}^{\mathsf{T}}, 
\label{}
\end{align}
we can obtain the mean vector and covariance matrix of the outputs after passing through the estimated phase.

On the basis of the above results, the quantum Fisher information turns out to be
\begin{align}
\nonumber {\cal F} =& \frac{1}{2}{\rm{Tr}}\left\{ {{\partial _\varphi }{\bf{\Sigma }}{{\left[ {{\bf{\Sigma }}{{\left( {{\partial _\varphi }{\bf{\Sigma }}} \right)}^{ - 1}}{{\bf{\Sigma }}^{\mathsf{T}} } + \frac{1}{4}{\bf{\Omega }}{{\left( {{\partial _\varphi }{\bf{\Sigma }}} \right)}^{ - 1}}{{\bf{\Omega }}^{\mathsf{T}} }} \right]}^{ - 1}}} \right\}   \\
& + {\left( {{\partial _\varphi }{\bf{M}}} \right)^{\mathsf{T}} }{{\bf{\Sigma }}^{ - 1}}\left( {{\partial _\varphi }{\bf{M}}} \right),
\label{qfi}
\end{align}
where
\begin{align}
{\mathbf{M}} = {{\mathbf{H}}}{{\mathbf{M}}_{\varphi}},
\label{}
\end{align}
\begin{align}
{\mathbf{\Sigma}} = {\mathbf{H}^{}}{\mathbf{\Sigma} _{\varphi}^{}} {\mathbf{H}_{}^{\mathsf{T}}}, 
\label{}
\end{align}
${\partial _\varphi }{\bf{M}} = {{\partial {\bf{M}}} \mathord{\left/
		{\vphantom {{\partial {\bf{M}}} {\partial \varphi }}} \right.
		\kern-\nulldelimiterspace} {\partial \varphi }}$ and
${\partial _\varphi }{\bf{\Sigma }} = {{\partial {\bf{\Sigma }}} \mathord{\left/
		{\vphantom {{\partial {\bf{\Sigma }}} {\partial \varphi }}} \right.
		\kern-\nulldelimiterspace} {\partial \varphi }}$.
The specific forms of two transformation matrices are given by
\begin{align}
\mathbf{\Omega}  = \left[ {\begin{array}{*{20}{c}}
	0&1&0&0\\
	{ - 1}&0&0&0\\
	0&0&0&1\\
	0&0&{ - 1}&0
	\end{array}} \right]
\label{}
\end{align}
and
\begin{align}
{\mathbf{H}} = \frac{1}{2}\left[ {\begin{array}{*{20}{c}}
	1&i&0&0\\
	1&{ - i}&0&0\\
	0&0&1&i\\
	0&0&1&{ - i}
	\end{array}} \right].
\label{}
\end{align}

After straightforward calculation, we finally get the general expression for the quantum Fisher information of our scheme

\begin{align}
\nonumber {\cal F} =& \frac{1}{{4{e^{4r}}}} [ 8{{\left| \alpha  \right|}^2}{e^{2r}}\left( {1 + {e^{ - 4g}}} \right)\left( {{e^{4r + 4g}}{{\cos }^2}\delta + {{\sin }^2}\delta } \right)  \\
& + {( {1 + {e^{8r}}} )\left( {1 + \cosh 4g} \right) } ] - 1.	
	\label{}
\end{align}

\begin{figure*}[htbp]
	\centering
	\includegraphics[width=0.32\textwidth]{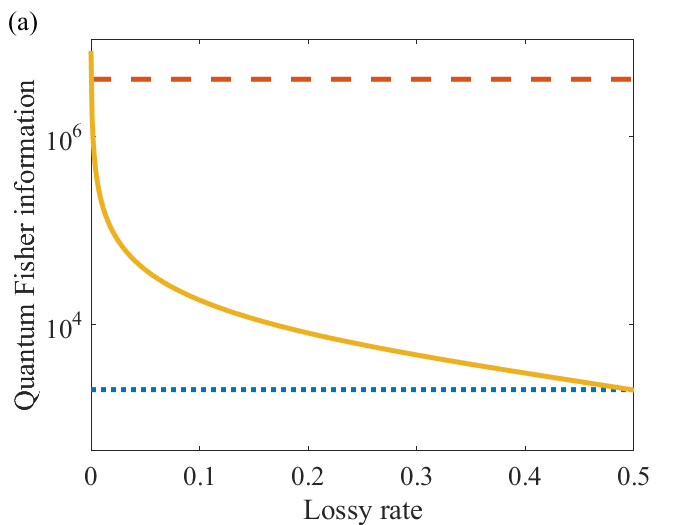}
   \includegraphics[width=0.32\textwidth]{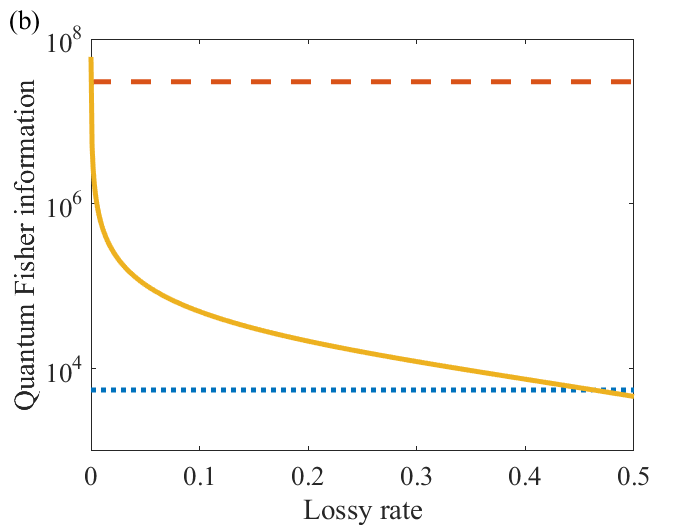}
   \includegraphics[width=0.32\textwidth]{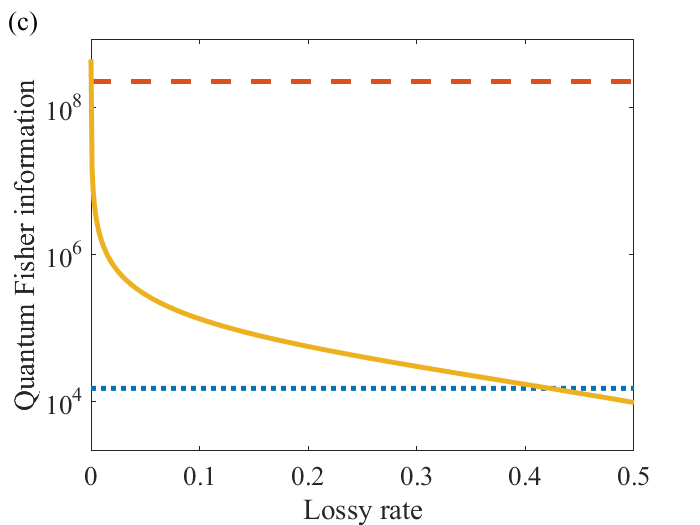}
	\caption{ The quantum Fisher information in the presence of photon loss (solid line) as function of lossy rate with $g =2.5$ and $N_{\rm{in}} = 2\sinh^2r$, (a) $r = 2$; (b) $r = 2.5$; (c) $r = 3$. Dotted line: the quantum Fisher information corresponding to the shot-noise limit (${\cal F}_{\rm {SNL}} = N$); dashed line: the quantum Fisher information corresponding to the Heisenberg limit (${\cal F}_{\rm {HL}} = N^2$).}
	\label{f2}
\end{figure*}

\begin{figure*}[htbp]
	\centering
	\includegraphics[width=0.32\textwidth]{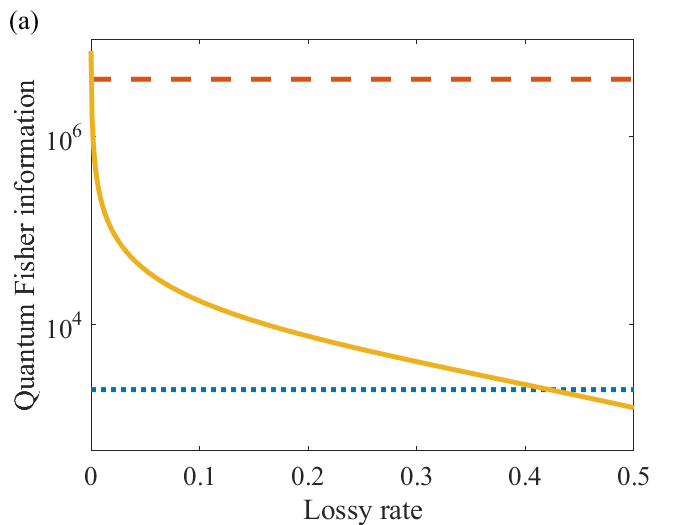}
   \includegraphics[width=0.32\textwidth]{rg25}
   \includegraphics[width=0.32\textwidth]{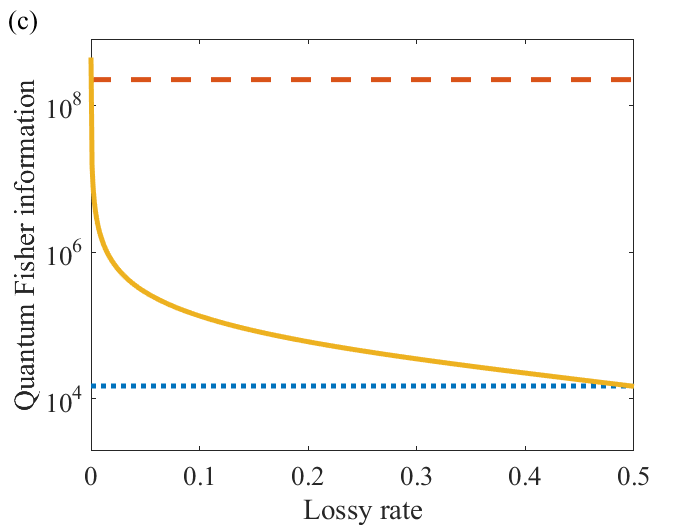}
	\caption{ The quantum Fisher information in the presence of photon loss (solid line) as a function of lossy rate with $r =2.5$ and $N_{\rm{in}} = 2\sinh^2r$, (a) $g = 2$; (b) $g = 2.5$; (c) $g = 3$. Dotted line: the quantum Fisher information corresponding to the shot-noise limit (${\cal F}_{\rm {SNL}} = N$); dashed line: the quantum Fisher information corresponding to the Heisenberg limit (${\cal F}_{\rm {HL}} = N^2$).}
	\label{f3}
\end{figure*}

\section{Optimal inputs and precision bound }
\label{III}

Given the current experimental techniques, we can reasonably assume 
${\left| \alpha  \right|^{\rm{2}}} \gg {\rm{1}}$,
${\sinh ^2}r \gg {\rm{1}}$ and
${\sinh ^2}g \gg {\rm{1}}$.
For simplicity, throughout this paper we use the following abbreviations
\begin{align}
G \equiv {\sinh ^2}g \approx {\cosh ^2}g \approx \frac{{{e^{2g}}}}{4} \gg {\rm{1}},
	\label{}
\end{align}

\begin{align}
S \equiv {\sinh ^2}r \approx \frac{{{e^{2r}}}}{4} \gg {\rm{1}}.	
	\label{}
\end{align}
We define the ratio of displacement portion to squeezing portion in the inputs as $k$, i.e.,
\begin{align}
k \equiv \frac{\left| \alpha  \right|^{\rm{2}}}{
{\sinh ^2}r}.
	\label{}
\end{align}
Related to this, the total average photon number can be approximately expressed as
\begin{align}
N \approx 4GS( {1 + 2k{{\cos }^2}\delta } )
	\label{}
\end{align}
and the quantum Fisher information can be approximately expressed as
\begin{align}
{\cal F} \approx 32G^2S^2( {1 + 4k{{\cos }^2}\delta } ).
	\label{}
\end{align}

It is not difficult to find the following inequality
\begin{align}
{1 + 4k{{\cos }^2}\delta } \le ({1 + 2k{{\cos }^2}\delta } )^2
	\label{}
\end{align}
with equality if the condition
$k = 0$ is satisfied.
Hence, $k = 0$ is the optimal ratio for maximizing the quantum Fisher information, and the optimal inputs are two squeezed vacuum states.
At this point, the average photon number of the inputs is $N_{\rm{in}} = 2\sinh^2r$ and the total average photon number is
\begin{align}
N = {N_{{\rm{in}}}}\cosh {\rm{2}}g  + {\rm{2}}{\sinh ^{\rm{2}}}g.
	\label{}
\end{align}
The corresponding quantum Fisher information is reduced to
\begin{align}
{{\cal F}_{{\rm{S}} \otimes {\rm{S}}}} = \frac{1}{{4{e^{4r}}}}\left( {1 + {e^{8r}} + 2{e^{4r}}\cosh 4g\cosh 4r} \right) - 1,	
	\label{}
\end{align}
and we have
\begin{align}
{{\cal F}_{{\rm{S}} \otimes {\rm{S}}}} \approx {\rm{32}}{G^2}{S^2} \approx 2{N^2}	
	\label{}
\end{align}
for large squeezing parameter and gain.
The above result indicates a sub-Heisenberg-limited precision bound, regardless of the ratio of average photon number of the inputs to total average photon number.

Since two squeezed vacuum states are used as the inputs, our scheme is a pure quantum scheme. 
In a lossless environment, it is superior to a pure classical scheme using two coherent states ($\approx 4N^2/3$) \cite{PhysRevA.93.023810} and a classical-quantum hybrid scheme using coherent along with squeezed vacuum states ($\approx 3N^2/2$) \cite{PhysRevA.94.063840}.

\section{Precision bound in the presence of photon loss}
\label{IV}

For any optical system, photon loss is always inevitable and is main hindrance to achieve high precision.
This process can be simulated by adding fictitious beam splitters.
The reflection of the fictitious beam splitter leads to a decrease in the number of photons; meanwhile, the coupling of vacuum fluctuation reduces the coherence of quantum states.
Both of these two factors result in a degradation of the precision.
For practical purposes, in this section we analyze the precision bound of our scheme in a lossy environment.

Generally, a quantum state after a lossy process becomes a mixed state.
At this point, it is quite difficult to give an analytical expression for the quantum Fisher information.
In particular, for Gaussian states, symplectic geometry method can provide an analytical result. 
Let us use $L$ to represent the lossy rate, then the mean vector and covariance matrix in a lossy environment can be written as \cite{PhysRevA.81.033819,gard2017nearly}
\begin{align}
{\mathbf{M}_{\varphi}} = \sqrt {1 - L} \cdot {{\mathbf{U}}_{{\rm{PE}}}}{{\mathbf{U}}_{{\rm{OPA}}}}{{\mathbf{M}}_{{\rm{in}}}},
\label{}
\end{align}
\begin{align}
{\mathbf{\Sigma}_{\varphi}} = (1 - L) {\mathbf{U}}_{\rm{PE}}^{}{\mathbf{U}}_{\rm{OPA}}^{}\Sigma _{\rm{in}}^{}{\mathbf{U}}_{\rm{OPA}}^{\mathsf{T}} {\mathbf{U}}_{\rm{PE}}^{\mathsf{T}}  + L {\mathbf{I}_{4}}.
\label{}
\end{align}
The above results suggest that a quantum state after photon loss becomes a statistical mixture of the quantum state and a vacuum state.

By substituting the above results into Eq. (\ref{qfi}), the quantum Fisher information in a lossy environment is found to be
\begin{align}
{{\cal F}_{{\rm{S}} \otimes {\rm{S}}}^{\rm L}} = \frac{{{{\left( {1 - L} \right)}^2}{\Delta _1}}}{{4{\Delta _2}}}
	\label{}
\end{align}
with
\begin{align}
\nonumber {\Delta _1} =& 4{e^{4r}}( {L - L^2} )\cosh 2g\cosh 2r\left[ {\cosh 4g + 2\cosh 4r - 3} \right] \\
\nonumber &+ ( {1 + {e^{8r}}} )( {2 - 2L + {L^2}} ) - 2{e^{4r}} ( {4 - 4L + 3{L^2}} ) \\	
&+ 2{e^{4r}} \cosh 4g[ {( {2 - 2L + 3{L^2}} )\cosh 4r - {L^2}} ]  
	\label{}
\end{align}
and
\begin{align}
\nonumber {\Delta _2} =& {e^{4r}}{L^2}[ {{{( {1 - L} )}^2}\cosh 4g + 2( {2 - 2L + {L^2}} ){{\cosh }^2}2r} ]   \\
\nonumber &+ 2{e^{2r}}\left( {1 + {e^{4r}}} \right)\left( {1 - L} \right)L\left( {1 - L + {L^2}} \right)\cosh 2g \\
&+ {e^{4r}}(2 - 4L + 5{L^2}).
\label{}
\end{align}

\begin{figure*}[htbp]
	\centering	\includegraphics[width=0.4\textwidth]{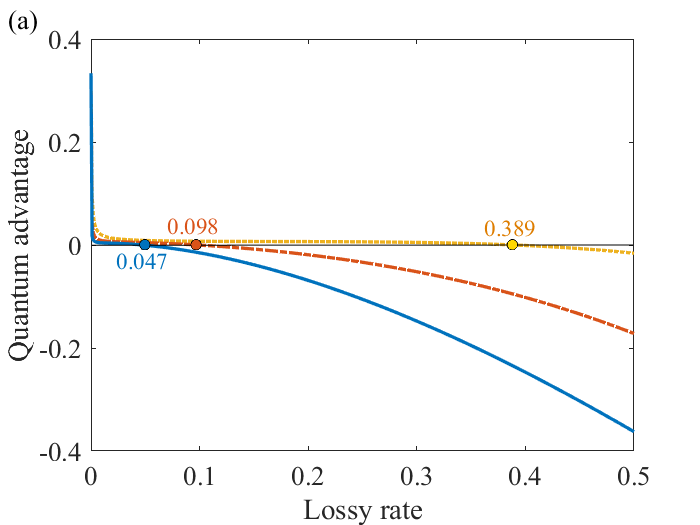}
    \includegraphics[width=0.4\textwidth]{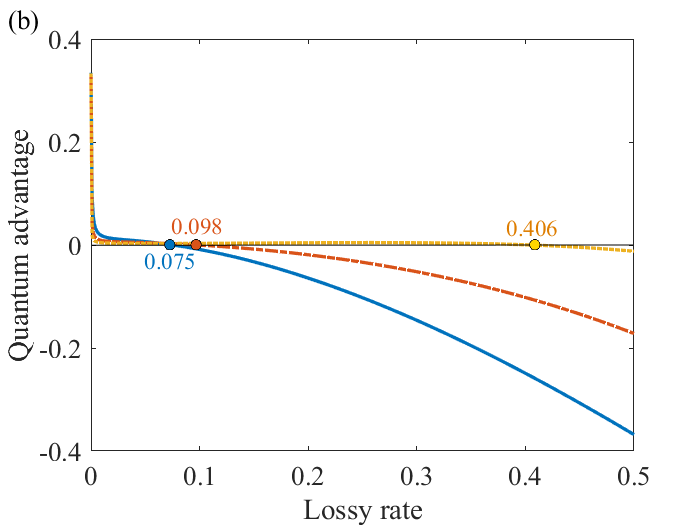}
	\caption{ The quantum advantage as a function of lossy rate. (a) $r =2.5$ and $g = 2$ (solid line); $g = 2.5$ (dash-dotted line); $g = 3$ (dotted line). (b) $g =2.5$ and $r = 2$ (dotted line); $r = 2.5$ (dash-dotted line); $r = 3$ (solid line). }
	\label{f4}
\end{figure*}

In Fig. \ref{f2} and Fig. \ref{f3} we give the quantum Fisher information against lossy rate to intuitively show the precision bound in the presence of photon loss.
It turns out that the precision bound is inferior to the Heisenberg limit even for a slight lossy rate.
As the lossy rate increases, the precision bound gradually degrades to the shot-noise limit.
For a fixed gain, the degradation of precision bound slightly slows down with the decrease of photon number of the inputs, as shown in Fig. \ref{f2}.
For a fixed photon number of the inputs, with the decrease of gain, the precision bound degrades to the shot-noise limit with a slightly faster pace, as shown in Fig. \ref{f3}.
This may be due to the fact that, for a fixed photon number of the inputs, the correlation between the two modes can be improved with increasing gain.
On the whole, it is beneficial for improving the tolerance against photon loss by reducing the proportion of photon number of the inputs when the total number of photons is fixed.

In general, pure quantum schemes are sensitive to photon loss. 
Here we compare the precision bound of our (pure quantum) scheme with that of the (classical-quantum hybrid) scheme using coherent along with squeezed vacuum states (see Appendix for detailed calculation).
Let us consider quantum advantage, which is defined as
\begin{align}
{{\cal A}_{\rm{Q}}}  = \frac{{{\cal F}_{{\rm{S}} \otimes {\rm{S}}}^{\rm L}}}{{{\cal F}_{{\rm{C}} \otimes {\rm{S}}}^{\rm L}}} - 1.
\label{}
\end{align}
The positive and negative values indicate that a pure quantum scheme is superior and inferior to a classical-quantum hybrid scheme, respectively

Figure \ref{f4} gives the dependence of quantum advantage on the lossy rate.
The advantage of the pure quantum scheme is remarkable for an extremely low lossy rate, but this advantage quickly disappears with the increase of the lossy rate. 
As the lossy rate further increases, the classical-quantum hybrid scheme becomes an optimal candidate.
In addition, the range of quantum advantages is larger for a high gain or a low photon number of the inputs.

\section{Conclusion}
\label{V}
In summary, we addressed the problem of quantum phase estimation using two-mode squeezed Gaussian states.
We analyzed the precision bound through the use of the quantum Fisher information.
By maximizing the precision bound, two squeezed vacuum states were determined as the optimal inputs.
For a lossless environment, the precision bound can outperform the Heisenberg limit by a factor of 2.
In the presence of photon loss, sub-shot-noise-limited precision bound can be attainable with the lossy rate below 0.4.
These results may be beneficial for practical quantum metrology based on nonlinear dynamics.

\section*{Acknowledgment} 
This work was supported by the National Natural Science Foundation of China (12104193) and the Program of Zhongwu Young Innovative Talents of Jiangsu University of Technology (20230013).

\appendix

\section*{Appendix} 
Here we provide the calculation of the quantum Fisher information of a lossy scheme using coherent and squeezed vacuum states.
We assume that the coherent and squeezed vacuum states are injected via modes $a$ and $b$, respectively.
The phase of displacement parameter and that of squeezing parameter are 0 and $\pi$, which are the optimal phase matching condition.
Accordingly, the mean vector of the inputs is given by 
\begin{align}
{\mathbf{M}'_{\rm{in}}} = [ {\begin{array}{*{20}{c}}
	{2 | \alpha | }&{0}&{0}&{0}
	\end{array}} ]^{\mathsf{T}},
\label{}
\end{align}
and the covariance matrix of the inputs is given by
\begin{align}
{\mathbf{\Sigma}' _{{\rm{in}}}} = \left[ {\begin{array}{*{20}{c}}
	{{1}}&0&0&0\\
	0&{{1}}&0&0\\
	0&0&{{e^{2r}}}&0\\
	0&0&0&{{e^{ - 2r}}}
	\end{array}} \right].
\label{}
\end{align}

Based on the method in the main text, we get the quantum Fisher
information
\begin{align}
{{\cal F}_{{\rm{C}} \otimes {\rm{S}}}^{\rm L}} = \frac{{1 - L}}{4}\left[ {\frac{{{\gamma _1}}}{{{\gamma _2}}} + \frac{{{\gamma _3}}}{{{\gamma _4}}}} \right]
\label{}
\end{align}	

with
\begin{widetext}
\begin{align}
\nonumber {\gamma _1} =&  4( {1 - L} ){\cosh ^2}r{\sinh ^2}2g[ { {L^2} + {e^{4r}}{L^2} + 2{e^{2r}}( {2 - 2L + {L^2}} ) + {( {1 + {e^{2r}}} )^2}( {1 - L})L\cosh 2g} ] \\
\nonumber & -4( {1 - L} ){\cosh ^2}r{\cosh ^2}g\{ {e^{4r}}( {8 - 17L + 15{L^2}} ) -{e^{2r}}( {24 - 38L + 34{L^2}} ) - {( {1 + {e^{2r}}} )^2}( {1 - L} )L\cosh 4g \\
& + 8 - 17L + 15{L^2} - 2\cosh 2g[ {4 - 7L + 8{L^2} + {e^{4r}}( {4 - 7L + 8{L^2}} ) - 2{e^{2r}}( {2 - 7L - 8{L^2}} )} ]  \}, 
\label{}
\end{align}

\begin{align}
\nonumber {\gamma _2} =&  {( {1 + {e^{2r}}} )^2}[ {4L( {1 - 2L + 2{L^2} - {L^3}} )\cosh 2g + {{( {1 - L} )}^2}{L^2}\cosh 4g} ] +{e^{2r}}( {8 - 16L + 18{L^2} - 12{L^3} + 6{L^4}} ) \\
&+ ( {1 + {e^{4r}}} )( {5{L^2} - 6{L^3} + 3{L^4}} ),
\label{}
\end{align}

\begin{align}
{\gamma _3} = 8{\left| \alpha  \right|^2}{\cosh ^2}g\left[ {1 - e^{2r}\left( {1 - 3L} \right) -L + (1+{e^{2r}})\left( {1 - L} \right){\cosh }2g} \right],
\label{}
\end{align}

\begin{align}
{\gamma _4} = 1-2L+\left( {1 + {e^{2r}}} \right) {L^2} + \left( {1 + {e^{2r}}} \right)\left( {1 - L} \right)L\cosh 2g.
\label{}
\end{align}

\end{widetext}


%

\end{document}